\documentclass[twocolumn,english,aps,pra]{revtex4-1}
\usepackage{color}
\usepackage{babel}
\usepackage{amsmath}
\usepackage{amssymb}
\usepackage{graphicx}
\usepackage[unicode=true,pdfusetitle,
 bookmarks=true,bookmarksnumbered=false,bookmarksopen=false,
 breaklinks=false,pdfborder={0 0 0},pdfborderstyle={},backref=false,colorlinks=true]
 {hyperref}
\hypersetup{
 citecolor=cyan,urlcolor=magenta,linkcolor=blue}

\makeatletter

\usepackage{amsthm}\usepackage{epstopdf}\usepackage{color}
\usepackage{bm}\usepackage{babel}

\makeatother

\begin{document}

\title{Identifying Orbital Angular Momentum of Vectorial Vortices with Pancharatnam
Phase and Stokes Parameters}

\author{Dengke Zhang}

\author{Xue Feng}
\email{x-feng@tsinghua.edu.cn}

\author{Kaiyu Cui}

\author{Fang Liu}

\author{Yidong Huang}

\affiliation{Department of Electronic Engineering, Tsinghua National Laboratory
for Information Science and Technology, Tsinghua University, Beijing
100084, China.}
\begin{abstract}
In this work, an explicit formula is deduced for identifying the orbital
angular moment (OAM) of vectorial vortex with space-variant state
of polarization (SOP). Different to scalar vortex, the OAM of vectorial
vortex can be attributed to two parts: the azimuthal gradient of Pancharatnam
phase and the product of the azimuthal gradient of orientation angle
of SOP and relevant solid angle on the Poincar\'{e} sphere. With our
formula, a geometrical description for OAM of light beams can be achieved
under the framework of the traditional Poincar\'{e} sphere. Numerical
simulations for two types of vectorial vortices have been carried
on to confirm our presented formula and demonstrate the geometrical
description of OAM. Furthermore, the finding will pave the way for
precise characterization of OAM charge of vectorial vortices.
\end{abstract}
\maketitle

\section{Introduction}

It is well known that light carries both linear and angular momenta
while the angular momenta (AM) can be divided into spin angular momentum
(SAM) and orbital angular momentum (OAM) \cite{Padgett2004PT,Franke-Arnold2008LPR,Barnett2002JOBQSO}.
Generally, in the paraxial approximation, it is believed that SAM
and OAM are associated with polarization and spatial profile of the
light fields, respectively \cite{Yao2011AOP}. As explicated by Allen
\textit{et~al.} in 1992 \cite{Allen1992PRA}, a scalar vortex field
with wavefront of $\exp(-il\phi)$ holds discrete OAM of $l\hbar$ per
photon, where $l$ is the topological charge. Thus, for scalar vortices,
the topological charge is directly related to the OAM of light beam.
However, for vectorial vortex fields, even in the paraxial approximation,
only the helical wavefront is insufficient to characterize OAM just
by utilizing topological charge while the state of polarization (SOP)
of light field should also be taken into account \cite{Zambrini2007OE,Freund2002OL}.
As demonstrated by Wang \textit{et~al.} in 2010 \cite{Wang2010PRL},
besides the azimuthal phase gradient, the OAM also can be generated
from the curl of polarization in a vectorial vortex field. Meanwhile,
Hasman \textit{et~al.} declared that there is a link between OAM
and geometric phase induced by space-variant SOP of light fields \cite{Bomzon2002OLa,Bomzon2002OL,Niv2006OE}.
But so far, the explicit relation between OAM and phase distribution
in vectorial vortex fields is still veiled.

In this work, we found that, for vectorial vortex, the OAM can be
attributed to two parts: the azimuthal gradient of Pancharatnam phase
and the product of the azimuthal gradient of orientation angle of
SOP and the related solid angle on the Poincar\'{e} sphere. Numerical
simulations have been carried on vectorial vertices generated by superposition
of two scalar vortex fields and phased array antenna, and both of
them have confirmed such relation. Further, since our deduced formula
of OAM charge is expressed with normal Stokes parameters, the traditional
Poincar\'{e} sphere can be utilized to fully characterize both the
SAM and OAM. It indicates that geometrical description and characterization
of OAM can be achieved by adopting the basic Poincar\'{e} sphere,
which is different to the previous reports based on multiple high-order
Poincar\'{e} spheres \cite{Milione2012PRL,Milione2011PRL,Padgett1999OL}.
On the other hand, as measuring Stokes parameters is a standard measurement
of polarization state, it can be expected that such formula could
provide an effective and accurate method for identifying the OAM charge,
which is very urgent in practical application of OAM beams \cite{Tamburini2012NJP,Wang2012NP,Nicolas2014NP,Dudley2014OE,Milione2015OL,Milione2015JO}.
Meanwhile, because of the explicit expression between OAM and SOP,
we believe that this work would provide a new sight of studies on
the vectorial vortices, spin-orbit interaction, and such related fields
\cite{Galvez2003PRL,Bliokh2006PRL,Aiello2009PRL,Bliokh2008PRL,Karimi2010PRA,Bliokh2008NP}.

\section{Theoretical description}

Under the paraxial approximation, the electric field of a fully polarized
vectorial vortex beam with angular frequency $\omega$ propagating
along $z$ direction in free space can be written as \cite{Torres2011}
\begin{equation}
\vec{E}(x,y)=i\omega\left(\alpha\hat{x}+\beta\hat{y}+\frac{i}{k}\left(\frac{\partial\alpha}{\partial x}+\frac{\partial\beta}{\partial y}\right)\hat{z}\right)e^{ikz},\label{eq:E-field}
\end{equation}
where $\alpha$ and $\beta$ represent the complex amplitude of $x-$
and $y-$component of electric field, respectively. Obviously, such
a vectorial vortex beam has space-variant SOP and its $z$-component
of angular momentum density can be calculated and divided into spin
and orbital parts in cylindrical coordinate system as 
\begin{align}
j_{z}^{\mathrm{spin}} & =i\omega\varepsilon_{0}r\frac{\partial}{\partial r}\left(\alpha^{\ast}\beta-\beta^{\ast}\alpha\right),\label{eq:j-spin}\\
j_{z}^{\mathrm{orbit}} & =i\frac{\omega\varepsilon_{0}}{2}\left(\alpha\frac{\partial}{\partial\phi}\alpha^{\ast}+\beta\frac{\partial}{\partial\phi}\beta^{\ast}-\alpha^{\ast}\frac{\partial}{\partial\phi}\alpha-\beta^{\ast}\frac{\partial}{\partial\phi}\beta\right),\label{eq:j-orbit}
\end{align}

As demonstrated in Ref.~\cite{Milione2012PRL}, an effective tool
for studying SOP of light is the Poincar\'{e} sphere with Stokes parameters.
Here, Stokes parameters and the Poincar\'{e} sphere are also introduced
to deduce the relation between OAM and SOP. In equations (\ref{eq:j-spin})
and (\ref{eq:j-orbit}), the complex amplitudes of $\alpha$ and $\beta$
can be written as $A_{x(y)}(x,y)e^{-i\delta_{x(y)}(x,y)}$, where
$A_{x(y)}$ and $\delta_{x(y)}$ are amplitude and phase (both are
real numbers), respectively. Then, the Stokes parameters would be
defined as \cite{Born1999}

\begin{equation}
\begin{aligned}S_{0} & =\tilde{A}_{x}^{2}+\tilde{A}_{y}^{2}\\
S_{1} & =\tilde{A}_{x}^{2}-\tilde{A}_{y}^{2}\\
S_{2} & =2\tilde{A}_{x}\tilde{A}_{y}\cos\delta_{s}\\
S_{3} & =2\tilde{A}_{x}\tilde{A}_{y}\sin\delta_{s}
\end{aligned}
\label{eq:Stokes}
\end{equation}

\noindent where $\tilde{A}_{x(y)}=A_{x(y)}/\sqrt{I_{\mathrm{E}}}$
are normalized to the electric intensity of $I_{\mathrm{E}}=A_{x}^{2}+A_{y}^{2}$
and $\delta_{s}=\delta_{y}-\delta_{x}$ is the phase difference between
$x-$ and $y-$components. Then using $S_{1}$, $S_{2}$, and $S_{3}$
as the sphere's Cartesian coordinates, the Poincar\'{e} sphere is
constructed and its corresponding orientation angle $\ensuremath{\psi}_{\mathrm{S}}$
of SOP on the Poincar\'{e} sphere can be resolved by 
\begin{equation}
\tan(2\psi_{\mathrm{S}})=S_{2}/S_{1}.\label{eq:orient-angle}
\end{equation}

With the ratio of angular momentum to energy examined by Allen \cite{Allen2000OC},
the average SAM charge and OAM charge of a vortex beam can be calculated.
The SAM charge can be solved by calculating the SAM density with $S_{3}$,
which directly represents the polarization degree \cite{Born1999}.
While for OAM charge, there is no explicit connection with Stokes
parameters. According to the feature of space-variant SOP in vectorial
vortex fields, Pancharatnam phase is adopted to reveal the phase distribution
for a vectorial vortex beam as shown in Ref.~\cite{Niv2006OE}. The
reason is that the Pancharatnam phase can well describe the phase
difference of lights with different SOP and the OAM is a quantity
related to the phase distribution of lights. As described in Ref.~\cite{Berry1987JMO},
Pancharatnam phase is defined as $\psi_{\mathrm{P}}=\arg(\left\langle \Phi_{\mathrm{A}}\mid\Phi_{\mathrm{B}}\right\rangle )$
between two different SOPs of $\left|\Phi_{\mathrm{A}}\right\rangle $
and $\left|\Phi_{\mathrm{B}}\right\rangle $. Based on mode expansion
theory, any optical beam can be expanded by right and left circularly
polarized light, which are written as $\left|\Phi_{\mathrm{R(L)}}\right\rangle =(\hat{x}\pm i\hat{y})/\sqrt{2}$.
For the same reason, in the paper, the right or the left circularly
polarized field is set as a reference field. Then the Pancharatnam
phase of the investigated vectorial vortex field $\left|\Phi_{\mathrm{E}}\right\rangle =\alpha\hat{x}+\beta\hat{y}$
(defined by equation (\ref{eq:E-field})) to the reference field is
given by 
\begin{equation}
\psi_{\mathrm{PR(L)}}=\arg(\left\langle \Phi_{\mathrm{R(L)}}\mid\Phi_{\mathrm{E}}\right\rangle ).\label{eq:Pan-phase}
\end{equation}

After some derivations (detailed in Appendix A), by applying the orientation
angle $\psi_{\mathrm{S}}$ of SOP on the Poincar\'{e} sphere and the
Pancharatnam phase $\psi_{\mathrm{PR(L)}}$ defined by equations (\ref{eq:orient-angle})
and (\ref{eq:Pan-phase}), the average OAM charge can be resolved
as 
\begin{equation}
l=\frac{\iint I_{\mathrm{E}}\left(-S_{0}\frac{\partial\psi_{\mathrm{PR(L)}}}{\partial\phi}\mp(S_{0}\pm S_{3})\frac{\partial\psi_{\mathrm{S}}}{\partial\phi}\right)rdrd\phi}{\iint I_{\mathrm{E}}S_{0}rdrd\phi}.\label{eq:OAM-charge}
\end{equation}

In bracket of numerator of equation (\ref{eq:OAM-charge}), the first
term is the derivative of spiral spatial phase, which is the topological
Pancharatnam charge similar to definition in Ref.~\cite{Niv2006OE},
and could be understood as the counterpart of topological charge in
scalar vortex fields. The second term is related to the variation
of SOP in space, which could be analyzed with Poincar\'{e} sphere.
To illustrate the physical interpretations and applicable scope of
equation (\ref{eq:OAM-charge}), in the following section, two cases
are demonstrated, where vectorial vertices are generated by superposition
of two scalar vortex fields and phased array antenna.

\section{Simulation results}

\begin{figure}
\includegraphics[width=8cm]{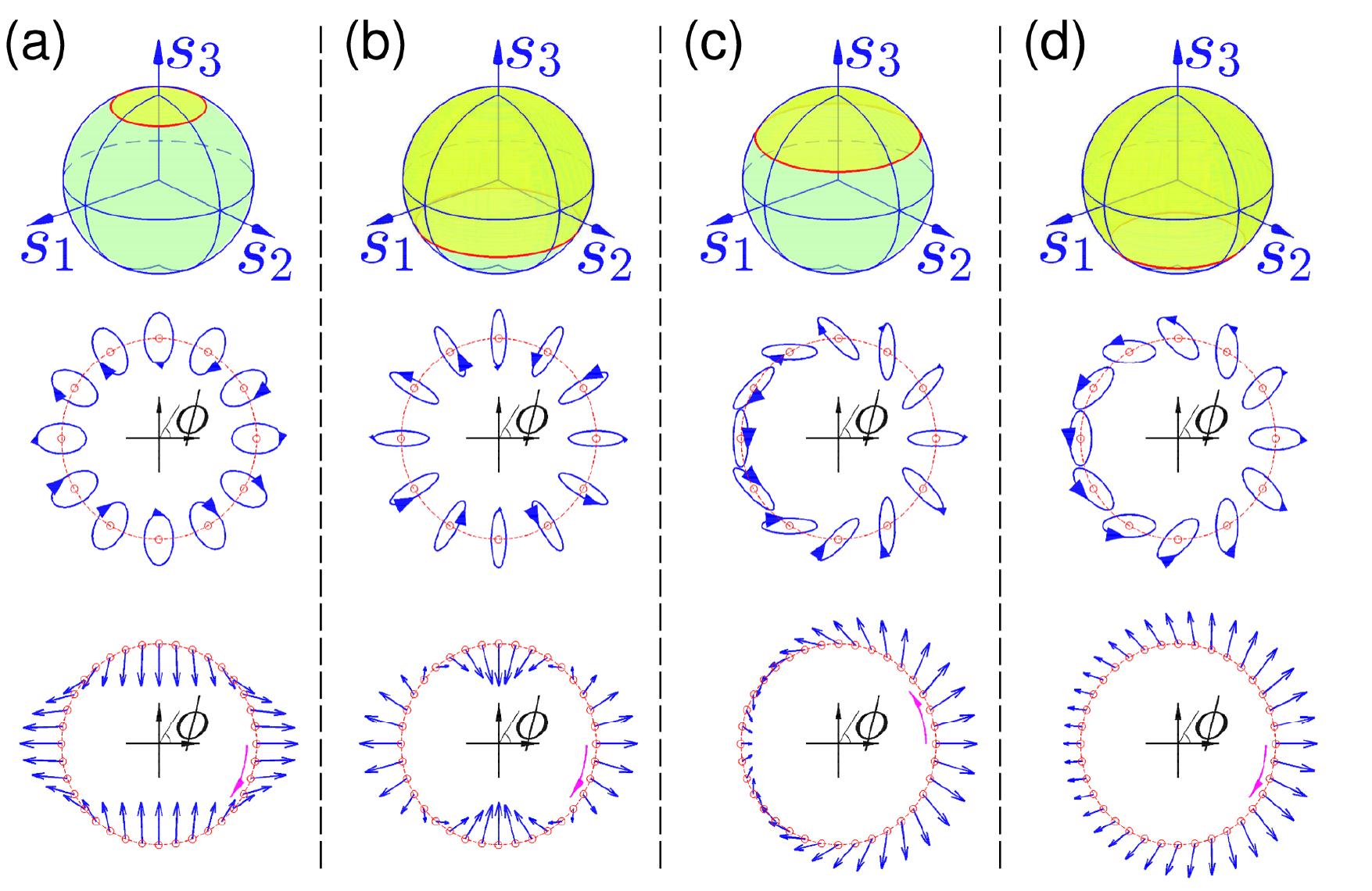} \caption{\textbf{SOP distributions.} Four vector beams with azimuthal variant
state of polarization (SOP) generated with equation (\ref{eq:superE-filed})
are shown in four row panels, corresponding to (a-d). In each panel,
different sketches of SOP trace on the Poincar\'{e} sphere marked
by red line, SOP distribution in space and snap picture of SOP are
demonstrated in order. Associated parameters in equation (\ref{eq:superE-filed})
for field generation are (a) $\{l_{\mathrm{L}},l_{\mathrm{R}}\}=\{1,3\}$,
$\theta=30^{\circ}$, (b) $\{l_{\mathrm{L}},l_{\mathrm{R}}\}=\{1,3\}$,
$\theta=120^{\circ}$, (c) $\{l_{\mathrm{L}},l_{\mathrm{R}}\}=\{-2,1\}$,
$\theta=60^{\circ}$, and (d) $\{l_{\mathrm{L}},l_{\mathrm{R}}\}=\{-2,1\}$,
$\theta=135^{\circ}$, respectively.}
\label{fig1} 
\end{figure}

\noindent \textit{\textbf{Superposition of two scalar vortex fields.}} For general vector beams, such as radially and azimuthally polarized
light, the field can be generated according to \cite{Maurer2007NJP}

\begin{align}
\left|\Phi_{\mathrm{E}}\right\rangle = & \frac{1}{\sqrt{2}}\cos\left(\frac{\theta}{2}\right)(\hat{x}-i\hat{y})e^{-il_{\mathrm{L}}\phi}\nonumber \\
 & +\frac{1}{\sqrt{2}}\sin\left(\frac{\theta}{2}\right)(\hat{x}+i\hat{y})e^{-il_{\mathrm{R}}\phi},\label{eq:superE-filed}
\end{align}

\noindent where $\theta$ is zenith angle in spherical coordinate
(the Poincar\'{e} sphere), and the set $\{l_{\mathrm{L}},l_{\mathrm{R}}\}$
is topological charge of field components with left and right circular
polarization respectively. For a fully polarized light ($S_{0}=1$),
there is a relation of $\Omega_{\mathrm{R(L)}}=2\pi(S_{0}\pm S_{3})$,
where $\Omega_{\mathrm{R(L)}}$ is the solid angle formed by the swept
surface area of SOP revolving around the south (north) pole on the
Poincar\'{e} sphere. Thus, equation (\ref{eq:OAM-charge}) can be
rewritten as 
\begin{equation}
l=\frac{\iint\left(-\frac{\partial\psi_{\mathrm{PR(L)}}}{\partial\phi}\mp\frac{\partial\psi_{\mathrm{S}}}{\partial\phi}\frac{\Omega_{\mathrm{R(L)}}}{2\pi}\right)rdrd\phi}{\iint{}rdrd\phi}.\label{eq:OAM-charge01}
\end{equation}
With field expression in equation (\ref{eq:superE-filed}), azimuthal
gradients of the Pancharatnam phase and the orientation angle could
be analytically expressed as (see Appendix B)

\begin{align}
\frac{\partial\psi_{\mathrm{PR(L)}}}{\partial\phi}=-l_{\mathrm{R(L)}},\label{eq:Pan-phase01}\\
\frac{\partial\psi_{\mathrm{S}}}{\partial\phi}=-\frac{l_{\mathrm{L}}-l_{\mathrm{R}}}{2}.\label{eq:orient-angle01}
\end{align}

Thus, substituting equations (\ref{eq:Pan-phase01}) and (\ref{eq:orient-angle01})
into equation (\ref{eq:OAM-charge01}), the OAM charge then read 
\begin{equation}
l=l_{\mathrm{R(L)}}\pm\frac{(l_{\mathrm{L}}-l_{\mathrm{R}})}{2}\frac{\Omega_{\mathrm{R(L)}}}{2\pi}.\label{eq:OAM-charge02}
\end{equation}
In the right side of equation (\ref{eq:OAM-charge02}), the first
term corresponds to the topological Pancharatnam charge ($l_{\mathrm{TPC}}$),
which is referenced to right or left circularly polarized field and
just equal to $l_{\mathrm{R(L)}}$ in this case. The second term is
the SOP-related charge, which is the product of the azimuthal gradient
of orientation angle of SOP and the related solid angle on the Poincar\'{e}
sphere. For more clarity, some simulations have been carried on four
fields generated with equation (\ref{eq:superE-filed}) and the results
are shown in Fig.~\ref{fig1}.

\begin{figure}
\includegraphics{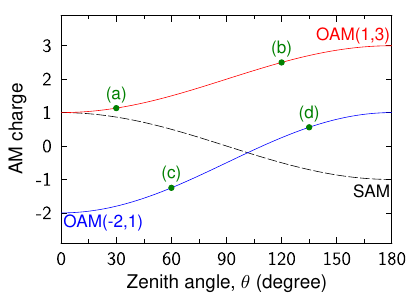} \caption{\textbf{Charges of vector beams generated by superposition of scalar
vortices.} The calculated OAM charge for the vector beams generated
with equation~(\ref{eq:superE-filed}) of $\{l_{\mathrm{L}},l_{\mathrm{R}}\}=\{1,3\}$
and $\{-2,1\}$ at different zenith angle in spherical coordinate
(the Poincar\'{e} sphere). Green dots of OAM charges are calculated
with our formula, which are corresponding to cases shown in Fig.~\ref{fig1}(a-d),
respectively. Solid lines are calculated by mode expansion method
according to equation~(\ref{eq:superE-filed}).}
\label{fig2} 
\end{figure}

Figure \ref{fig1}(a-d) are the calculated results while the parameters
are set as $\{l_{\mathrm{L}},l_{\mathrm{R}}\}=\{1,3\}$ with $\theta=30^{\circ},120^{\circ}$
and $\{l_{\mathrm{L}},l_{\mathrm{R}}\}=\{-2,1\}$ with $\theta=60^{\circ},135^{\circ}$.
For each row panel, there are three parts in order: SOP trace on Poincar\'{e}
sphere marked by red line, SOP distribution in space, and a SOP snap
in space. In Fig.~\ref{fig2}, calculated results of OAM charges
are shown as the green dots, which are calculated by equation (\ref{eq:OAM-charge02})
for the four cases shown in Fig.~\ref{fig1}(a-d). For comparison,
the OAM charges are also calculated by mode expansion method according
to equation (\ref{eq:superE-filed}) and shown as solid lines in Fig.~\ref{fig2}.
For the cases shown in Fig.~\ref{fig1}(a) and \ref{fig1}(b), the
left circularly polarized fields (north pole on the Poincar\'{e} sphere)
is selected as the reference field and swept surface areas are also
shown with yellow zone. While for Fig.~\ref{fig1}(c) and \ref{fig1}(d),
right circularly polarized field (south pole on the Poincar\'{e} sphere)
is selected as the reference. In Fig.~\ref{fig2}, all the calculated
results with our formula are in good agreement with those calculated
by mode expansion method. From the results shown in Figs.~\ref{fig1}
and \ref{fig2}, a clear relation of OAM charge versus the Pancharatnam
phase, orientation angle of SOP, and the related solid angle on the
Poincar\'{e} sphere is presented. Furthermore, with our formula, a
geometrical description of OAM can be obtained by utilizing a basic
Poincar\'{e} sphere, as shown in Fig.~\ref{fig1}.

\begin{figure}
\includegraphics{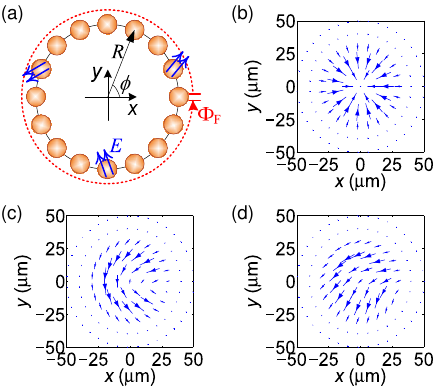} \caption{\textbf{Vectorial vortices generated with PAA.} (a) Schematic of the
considered phased array antenna (PAA), which consists of 16 units.
Each unit emits linearly polarized Gaussian beam and the polarization
direction and initial phase can be set. With PAA, varied vectorial
vortex beams can be generated, (b) state of polarization (SOP) makes
two revolutions at latitude on the Poincar\'{e} sphere (radially polarized
vectorial beam), (c) SOP makes one revolution (L-line vortex), and
(d) SOP makes half revolution.}
\label{fig3} 
\end{figure}

\begin{figure*}
\includegraphics{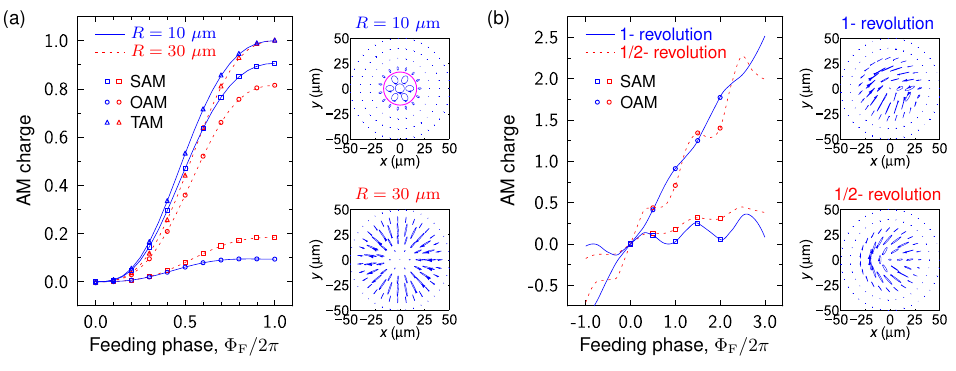} \caption{\textbf{Charges of vectorial vortices generated by PAA.} Calculated
angular momentum (AM) charges of (a) the vortex beam shown in Fig.~\ref{fig3}(b)
at different feeding phase of two different PAA radius of 10 and 30
$\mu$m, and (b) the vortex beam shown in Fig.~\ref{fig3}(c) and
\ref{fig3}(d) at different feeding phase of fixed PAA radius of 20
$\mu$m. In the figures, lines are calculated by Barnett's method
and symbols are calculated by our formula for OAM and Stoke parameter
of $S_{3}$ for SAM. Right-side insets also show the corresponding
SOP distribution at feed phase of $2\pi$.}
\label{fig4} 
\end{figure*}

\smallskip{}
\noindent \textit{\textbf{Phased array antenna.}} Recently, more and more attentions have been focused on the generation
of OAM beams with phased array antenna (PAA) in RF, microwave, and
lightwave region \cite{Mohammadi2010ITAP,Zhang2012OE,Cai2012S,Milione2011CLEO}.
To model such process, some simulations are also carried on an annular
PAA with antenna unit of linearly polarized Gaussian beam as schematically
shown in Fig.~\ref{fig3}(a). In the simulations, the optical communication
wavelength of 1550~nm is adopted. For each Gaussian beam, the waist
size is 8~$\mu$m and polarization direction is azimuthal-dependent.
The unit number is 16 and radius ($R$) of annular PAA, which is defined
by the distance between the PAA center and each unit center as marked
in Fig.~\ref{fig3}(a), can be adjusted. These parameters ensure
that the generated beam satisfies the paraxial approximation, thus
Barnett's method \cite{Barnett2002JOBQSO} can be utilized as a reference
with the results calculated by equation (\ref{eq:OAM-charge}). As
demonstrated in Ref.~\cite{Zhang2012OE}, the AM charges of the generated
beam can be tuned by varying the phase difference between adjacent
units. In our simulation, the adjacent phase differences are uniform
and the whole feeding phase of a circle ($\Phi_{\mathrm{F}}$) is
used to describe the setting phase of PAA. With such a structure,
various vortex beams can be generated, such as radially or azimuthally
polarized vector beams, L-line vortex beams \cite{Nye1983PRSLA},
and so on.

Figure \ref{fig3}(b) shows a radially polarized vectorial beam, where
SOP makes two revolutions at latitude on the Poincar\'{e} sphere for
a circle in the space. Corresponding AM charges are calculated by
both Barnett's method and our formula, which are shown as lines and
dots in Fig.~\ref{fig4}(a), respectively. Two cases with different
PAA radius of 10 and 30~$\mu$m are also considered under the varied
feeding phase, both methods give consistent OAM charge. Furthermore,
Fig.~\ref{fig3}(c) and \ref{fig3}(d) display another two types
of vortex beams, where SOP makes one and half revolution at latitude
on the Poincar\'{e} sphere for a circle in the space, respectively.
For a fixed PAA radius of 20~$\mu$m, the OAM charges at different
feeding phase are also calculated and presented in Fig.~\ref{fig4}(b),
and again, they are also in a very good agreement. These results indicate
that the calculations for OAM charge with equation (\ref{eq:OAM-charge})
can be applied on not only general vector beams but also complex vortex
beams under the paraxial approximation, which can be explained by
the principle of superposition with basis beams \cite{Goette2008OE}.

\section{Discussion}

It should be noticed that, for most general vector beam shown in Fig.~\ref{fig3}(b),
the feeding phase are transferred to both OAM and SAM (see Fig.~\ref{fig4}(a)),
which is quite different to the scalar vortex beam. For a scalar vortex
beam, the feeding phase $\Phi_{\mathrm{F}}$ would be fully transferred
to OAM. Even for the cases of $\Phi_{\mathrm{F}}=2N\pi$ ($N$ is
any integer number), the number of $N$ will be the value of total
angular momentum (TAM) charge of generated beam, while not the OAM
charge (recently, a similar report was presented in Ref.~\cite{Zhu2014OL}).
The reason is that some feeding phase is transferred to SAM in the
central zone of vortex as shown with magenta circle in right-side
inset of Fig.~\ref{fig4}(a) and meanwhile the reduction of solid
angle of swept area on Poincar\'{e} sphere would suppress the transformation
of OAM from feeding phase. Fortunately, through a carefully designed
PAA, the proportion of OAM charge can be varied by reducing power
proportion of field around vortex center. For the same reason of partial
OAM induced by azimuthal gradient of SOP-related phase, the detection
of OAM charge will be different with that for scalar vortex by only
detecting phase angle of wavefront. Thus, to characterize OAM charge
of vectorial vortices, new method is required. Here, we predict that
such detection can be achieved by traditional measurement of Stokes
parameters according to equation (\ref{eq:OAM-charge}).

In equation (\ref{eq:OAM-charge}), we introduce a reference field
to calculate the OAM charge of vortices. In this paper, only special
reference field, SOP of right- or left-handed circular polarization,
was adopted. However, this does not mean that a general reference
field would induce an incorrect calculation result of OAM charge.
To demonstrate this fact, a series of simulation were carried out
to make a contrast, which is explained in detail in Appendix C. Although
the selection of reference field do not affect the result of OAM charge,
right or left circularly polarized field is
a normal choice in the measurement of Stokes parameters and can be
induced a simple and elegant expression of OAM charge as equation
(\ref{eq:OAM-charge}).

In summary, for paraxial vectorial vortex beams propagating in free
space, it is deduced that the OAM charge is not only related with
the topological Pancharatnam charge but also the SOP-related charge
induced by space-variant state of polarization (SOP). Based on such
a connection, OAM also can be fully represented by the fundamental
Poincar\'{e} sphere. And we predict that the detection of OAM charge
can be achieved by testing Stokes parameters, which is a standard
test of polarization measurement for antennas. Moreover, because of
the explicit relation with SOP, we believe that this work would give
some new insights for studies on vectorial vortices, spin-orbit interaction,
photonic topological insulators, and so on.

\textit{Acknowledgements.} This work was supported by the National
Basic Research Program of China (Grant No. 2011CBA00608), the National
Natural Science Foundation of China (Grant No. 61307068 and 61321004)
and by the Opened Fund of the State Key Laboratory on Integrated Optoelectronics,
China. No. IOSKL2013KF09. The authors would like to thank Dr. Yu Wang,
Mr. Peng Zhao and Dr. Wei Zhang for their valuable discussions and
helpful comments.


%

\widetext 
\clearpage

\renewcommand{\theequation}{A\arabic{equation}} 
\setcounter{equation}{0} 
\renewcommand{\thefigure}{A\arabic{figure}} 
\setcounter{figure}{0}

\section*{Appendix A: Relation of orbital angular momentum, Pancharatnam phase,
and Stokes Parameters}

Under the paraxial approximation, the electric and magnetic fields
of a fully polarized vectorial vortex beam of angular frequency $\omega$
propagating along $z$ direction can be written as \cite{Torres2011} 

\begin{subequations} 
\begin{align}
\vec{E}\left(x,y\right) & =i\omega\left(\alpha\hat{x}+\beta\hat{y}+\frac{i}{k}\left(\frac{\partial\alpha}{\partial x}+\frac{\partial\beta}{\partial y}\right)\hat{z}\right)e^{ikz},\label{S1a}\\
\vec{B}\left(x,y\right) & =ik\left(-\beta\hat{x}+a\hat{y}+\frac{i}{k}\left(-\frac{\partial\beta}{\partial x}+\frac{\partial\alpha}{\partial y}\right)\hat{z}\right)e^{ikz},\label{S1b}
\end{align}
\end{subequations} 

\noindent where $\alpha$ and $\beta$ represent the complex amplitude
of of $x-$ and $y-$component of electric field. They can be written
as 

\begin{subequations} 
\begin{align}
\alpha(x,y) & =A_{x}(x,y)e^{-i\delta_{x}(x,y)},\label{S2a}\\
\beta(x,y) & =A_{y}(x,y)e^{-i\delta_{y}(x,y)},\label{S2b}
\end{align}
\end{subequations} 

\noindent where $A_{x\left(y\right)}$ and $\delta_{x\left(y\right)}$
are real numbers and represent amplitude and phase, respectively.
Thus, Stokes parameters are defined by \cite{Born1999}

\begin{equation}
\begin{aligned}S_{0}(x,y) & =\tilde{A}_{x}^{2}+\tilde{A}_{y}^{2},\\
S_{1}(x,y) & =\tilde{A}_{x}^{2}-\tilde{A}_{y}^{2},\\
S_{2}(x,y) & =2\tilde{A}_{x}\tilde{A}_{y}\cos\delta_{s},\\
S_{3}(x,y) & =2\tilde{A}_{x}\tilde{A}_{y}\sin\delta_{s}
\end{aligned}
\label{S3}
\end{equation}

\noindent where $\tilde{A}_{x\left(y\right)}=A_{x\left(y\right)}/\sqrt{I_{\text{E}}}$
are normalized electric field components with electric intensity of
$I_{\text{E}}=A_{x}^{2}+A_{y}^{2}$ and $\delta_{s}=\delta_{y}-\delta_{x}$
the phase difference between $x$ and $y$ electric field components.
Then using $S_{1}$, $S_{2}$, and $S_{3}$ as the sphere's Cartesian
coordinates, the Poincar\'{e} sphere is constructed and the corresponding
spherical angles $\left(2\psi_{\text{S}},2\chi_{\text{S}}\right)$
are resolved by \cite{Born1999} 

\noindent \begin{subequations} 
\begin{align}
\tan(2\psi_{\text{S}}) & =S_{2}/S_{1},\label{S4a}\\
\sin(2\chi_{\text{S}}) & =S_{3}/S_{0},\label{S4b}
\end{align}
\end{subequations}

The linear momentum density, which is defined as $\vec{p}=\varepsilon_{0}\vec{E}\times\vec{B}$,
can be written and divided into transverse and longitudinal components

\begin{subequations} 
\begin{align}
\vec{p}_{\bot} & =i\frac{\omega\varepsilon_{0}}{2}\left[\left(\alpha\nabla\alpha^{*}+\beta\nabla\beta^{*}-\alpha^{*}\nabla\alpha-\beta^{*}\nabla\beta\right)+2\nabla\times\left(\left(\alpha^{*}\beta-\beta^{*}\alpha\right)\hat{z}\right)\right],\label{S5a}\\
p_{z} & =\omega k\varepsilon_{0}\left(\left|\alpha\right|^{2}+\left|\beta\right|^{2}\right)=\omega k\varepsilon_{0}I_{\text{E}}S_{0}.\label{S5b}
\end{align}
\end{subequations} 

Meanwhile, the energy density of such a beam is 
\begin{equation}
w=cp_{z}=\varepsilon_{0}\omega^{2}\left(\left|\alpha\right|^{2}+\left|\beta\right|^{2}\right)=\varepsilon_{0}\omega^{2}I_{\text{E}}S_{0}.\label{S6}
\end{equation}
Then, the cross product of line momentum density with $\vec{r}$ (radius
vector) gives the angular momentum density, and $z-$component of
angular momentum density is 
\begin{equation}
\begin{split}j_{z} & =\left(\vec{r}\times\vec{p}\right)_{z}=rp_{\phi}\\
 & =i\frac{\omega\varepsilon_{0}}{2}\left[\left(\alpha\frac{\partial}{\partial\phi}\alpha^{*}+\beta\frac{\partial}{\partial\phi}\beta^{*}-\alpha^{*}\frac{\partial}{\partial\phi}\alpha-\beta^{*}\frac{\partial}{\partial\phi}\beta\right)+2r\frac{\partial}{\partial r}\left(\alpha^{*}\beta-\beta^{*}\alpha\right)\right].
\end{split}
\label{S7}
\end{equation}
Further, $j_{z}$ can be divided into spin and orbital parts as 

\begin{subequations} 
\begin{align}
j_{z}^{\text{spin}} & =i\omega\varepsilon_{0}r\frac{\partial}{\partial r}\left(\alpha^{*}\beta-\beta^{*}\alpha\right)=\omega\varepsilon_{0}r\frac{\partial\left(I_{\text{E}}S_{3}\right)}{\partial r},\label{S8a}\\
j_{z}^{\text{orbit}} & =i\frac{\omega\varepsilon_{0}}{2}\left(\alpha\frac{\partial}{\partial\phi}\alpha^{*}+\beta\frac{\partial}{\partial\phi}\beta^{*}-\alpha^{*}\frac{\partial}{\partial\phi}\alpha-\beta^{*}\frac{\partial}{\partial\phi}\beta\right)\nonumber \\
 & =\omega\varepsilon_{0}I_{\text{E}}\left(\tilde{A}_{x}^{2}\frac{\partial\delta_{x}}{\partial\phi}+\tilde{A}_{y}^{2}\frac{\partial\delta_{y}}{\partial\phi}\right).\label{S8b}
\end{align}
\end{subequations} 

With the ratio of angular momentum over energy that is examined by
Allen \cite{Allen2000OC}, the average SAM charge and OAM charge can
be calculated as

\begin{subequations}\label{S9} 
\begin{align}
s & =\omega\frac{\iint j_{z}^{\text{spin}}rdrd\phi}{\iint wrdrd\phi}=\frac{\iint I_{\text{E}}S_{3}rdrd\phi}{\iint I_{\text{E}}S_{0}rdrd\phi},\label{S9a}\\
l & =\omega\frac{\iint j_{z}^{\text{orbit}}rdrd\phi}{\iint wrdrd\phi}=\frac{\iint I_{\text{E}}\left(\tilde{A}_{x}^{2}\frac{\partial\delta_{x}}{\partial\phi}+\tilde{A}_{y}^{2}\frac{\partial\delta_{y}}{\partial\phi}\right)rdrd\phi}{\iint I_{\text{E}}S_{0}rdrd\phi}.\label{S9b}
\end{align}
\end{subequations} 

Then, introducing Pancharatnam phase for two different SOPs of $\left|\Phi_{\text{A}}\right\rangle $
and $\left|\Phi_{\text{B}}\right\rangle $, which is defined by \cite{Berry1987JMO}
\begin{equation}
\psi_{\text{P}}=\arg\left(\left\langle \Phi_{\text{A}}|\Phi_{\text{B}}\right\rangle \right).\label{S10}
\end{equation}
Here, using right or left circularly polarized field as reference
field, the Pancharatnam phase of any field $\left|\Phi_{\text{E}}\right\rangle =\alpha\hat{x}+\beta\hat{y}$
can be written as 
\begin{equation}
\psi_{\text{PR(L)}}=\arg\left(\left\langle \Phi_{\text{R(L)}}|\Phi_{\text{E}}\right\rangle \right).\label{S11}
\end{equation}
Then we can obtain 
\begin{equation}
\tan\psi_{\text{PR(L)}}=\frac{\tilde{A}_{y}\cos\delta_{y}\pm\tilde{A}_{x}\sin\delta_{x}}{\tilde{A}_{y}\sin\delta_{y}\mp\tilde{A}_{x}\cos\delta_{x}}.\label{S12}
\end{equation}
Further, we can deduce the azimuthal gradient of the Pancharatnam
phase as 

\begin{subequations} \label{S13} 
\begin{align}
\frac{\partial\psi_{\text{PR}}}{\partial\phi} & =\frac{-\cos\delta_{s}}{S_{0}-S_{3}}\left(\tilde{A}_{x}\frac{\partial\tilde{A}_{y}}{\partial\phi}-\tilde{A}_{y}\frac{\partial\tilde{A}_{x}}{\partial\phi}\right)+\frac{S_{3}}{2\left(S_{0}-S_{3}\right)}\left(\frac{\partial\delta_{x}}{\partial\phi}+\frac{\partial\delta_{y}}{\partial\phi}\right)-\frac{1}{S_{0}-S_{3}}\left(\tilde{A}_{x}^{2}\frac{\partial\delta_{x}}{\partial\phi}+\tilde{A}_{y}^{2}\frac{\partial\delta_{y}}{\partial\phi}\right),\label{S13a}\\
\frac{\partial\psi_{\text{PL}}}{\partial\phi} & =\frac{\cos\delta_{s}}{S_{0}+S_{3}}\left(\tilde{A}_{x}\frac{\partial\tilde{A}_{y}}{\partial\phi}-\tilde{A}_{y}\frac{\partial\tilde{A}_{x}}{\partial\phi}\right)-\frac{S_{3}}{2\left(S_{0}+S_{3}\right)}\left(\frac{\partial\delta_{x}}{\partial\phi}+\frac{\partial\delta_{y}}{\partial\phi}\right)-\frac{1}{S_{0}+S_{3}}\left(\tilde{A}_{x}^{2}\frac{\partial\delta_{x}}{\partial\phi}+\tilde{A}_{y}^{2}\frac{\partial\delta_{y}}{\partial\phi}\right).\label{S13b}
\end{align}
\end{subequations} 

On the other hand, using equations \eqref{S4a} and \eqref{S3}, we
can obtain 
\begin{equation}
\frac{\partial\psi_{\text{S}}}{\partial\phi}=\frac{S_{0}\cos\delta_{s}}{S_{0}^{2}-S_{3}^{2}}\left(\tilde{A}_{x}\frac{\partial\tilde{A}_{y}}{\partial\phi}-\tilde{A}_{y}\frac{\partial\tilde{A}_{x}}{\partial\phi}\right)-\frac{S_{1}S_{3}}{2\left(S_{0}^{2}-S_{3}^{2}\right)}\left(\frac{\partial\delta_{y}}{\partial\phi}-\frac{\partial\delta_{x}}{\partial\phi}\right).\label{S14}
\end{equation}
With relation of 
\[
S_{0}\left(\frac{\partial\delta_{x}}{\partial\phi}+\frac{\partial\delta_{y}}{\partial\phi}\right)-S_{1}\left(\frac{\partial\delta_{y}}{\partial\phi}-\frac{\partial\delta_{x}}{\partial\phi}\right)=2\left(\tilde{A}_{x}^{2}\frac{\partial\delta_{x}}{\partial\phi}+\tilde{A}_{y}^{2}\frac{\partial\delta_{y}}{\partial\phi}\right)
\]
and equations \eqref{S13a}, \eqref{S13b}, and \eqref{S14}, the
following equations are obtained 

\begin{subequations} \label{S15} 
\begin{align}
S_{0}\frac{\partial\psi_{\text{PR}}}{\partial\phi} & =-(S_{0}+S_{3})\frac{\partial\psi_{\text{S}}}{\partial\phi}-\left(\tilde{A}_{x}^{2}\frac{\partial\delta_{x}}{\partial\phi}+\tilde{A}_{y}^{2}\frac{\partial\delta_{y}}{\partial\phi}\right),\label{S15a}\\
S_{0}\frac{\partial\psi_{\text{PL}}}{\partial\phi} & =(S_{0}-S_{3})\frac{\partial\psi_{\text{S}}}{\partial\phi}-\left(\tilde{A}_{x}^{2}\frac{\partial\delta_{x}}{\partial\phi}+\tilde{A}_{y}^{2}\frac{\partial\delta_{y}}{\partial\phi}\right).\label{S15b}
\end{align}
\end{subequations} 

Through further derivation with equations \eqref{S15a} and \eqref{S15b},
we can finally obtain 
\begin{equation}
\tilde{A}_{x}^{2}\frac{\partial\delta_{x}}{\partial\phi}+\tilde{A}_{y}^{2}\frac{\partial\delta_{y}}{\partial\phi}=-S_{0}\frac{\partial\psi_{\text{PR}\left(\text{L}\right)}}{\partial\phi}\mp(S_{0}\pm S_{3})\frac{\partial\psi_{\text{S}}}{\partial\phi}.\label{S16}
\end{equation}
Then, substituting equation \eqref{S16} into equation \eqref{S9b},
we can solve the OAM charge. In equation \eqref{S16}, the first term
is the derivative of $\psi_{\text{PR}\left(\text{L}\right)}$ known
as the topological Pancharatnam charge, and the second term is the
SOP-related charge.

\section*{Appendix B: OAM charge of vector beams generated by two scalar vortex
beams}

General vector beam can be generated by mode expansion as \cite{Maurer2007NJP}
\begin{equation}
\left|\Phi_{\text{E}}\right\rangle =\frac{1}{\sqrt{2}}\cos\left(\frac{\theta}{2}\right)\left(\hat{x}-i\hat{y}\right)e^{-il_{\text{L}}\phi}+\frac{1}{\sqrt{2}}\sin\left(\frac{\theta}{2}\right)\left(\hat{x}+i\hat{y}\right)e^{-il_{\text{R}}\phi-i\varphi_{0}}.\label{S17}
\end{equation}
Thus, with equation \eqref{S11}, we can obtain 
\begin{equation}
\frac{\partial\psi_{\text{PR}\left(\text{L}\right)}}{\partial\phi}=-l_{\text{R(L)}}.\label{S18}
\end{equation}
Further, the Stokes parameters can be deduced 
\begin{equation}
\begin{split}S_{1} & =\sin\theta\cos\left(-\left(l_{\text{L}}-l_{\text{R}}\right)\phi-\varphi_{0}\right),\\
S_{2} & =\sin\theta\sin\left(-\left(l_{\text{L}}-l_{\text{R}}\right)\phi-\varphi_{0}\right).
\end{split}
\label{S19}
\end{equation}
Thus, with equations \eqref{S4a} and \eqref{S3}, we obtain 
\begin{equation}
\psi_{\text{S}}=-\left(\left(l_{\text{L}}-l_{\text{R}}\right)\phi+\varphi_{0}\right)/2.\label{S20}
\end{equation}
Then, 
\begin{equation}
\frac{\partial\psi_{\text{S}}}{\partial\phi}=-\left(l_{\text{L}}-l_{\text{R}}\right)/2.\label{S21}
\end{equation}
Substituting equations \eqref{S18} and \eqref{S21} into equations
\eqref{S9b} and \eqref{S16}, the OAM charge can be written as 

\begin{subequations} \label{S22} 
\begin{align}
l & =l_{\text{R}}+\frac{\left(l_{\text{L}}-l_{\text{R}}\right)}{2}\left(1+\sin2\chi_{\text{S}}\right)=l_{\text{TPC}}^{\text{R}}+\frac{\left(l_{\text{L}}-l_{\text{R}}\right)}{2}\frac{\Omega{}_{\text{R}}}{2\pi}=l_{\text{TPC}}^{\text{R}}+\frac{\phi_{\text{gO}}^{\text{R}}}{\pi},\label{S22a}\\
l & =l_{\text{L}}-\frac{\left(l_{\text{L}}-l_{\text{R}}\right)}{2}\left(1-\sin2\chi_{\text{S}}\right)=l_{\text{TPC}}^{\text{L}}-\frac{\left(l_{\text{L}}-l_{\text{R}}\right)}{2}\frac{\Omega_{\text{L}}}{2\pi}=l_{\text{TPC}}^{\text{L}}-\frac{\phi_{\text{gO}}^{\text{L}}}{\pi}.\label{S22b}
\end{align}
\end{subequations} 

\noindent where $l_{\text{TPC}}^{\text{R(L)}}=l_{\text{R(L)}}$ is
the topological Pancharatnam charge that is referenced to right (left)
circularly polarized field, $\Omega_{\text{R}\left(\text{L}\right)}$
is the solid angle formed by the swept surface area of SOP revolving
around the south (north) pole on the Poincar\'{e} sphere, and the
$\phi_{\text{gO}}^{\text{R}\left(\text{L}\right)}$ is the equivalent
SOP-related phase induced by space-variant SOP.

\section*{Appendix C: Reference field with general SOP}

\begin{figure*}[!htb]
\includegraphics{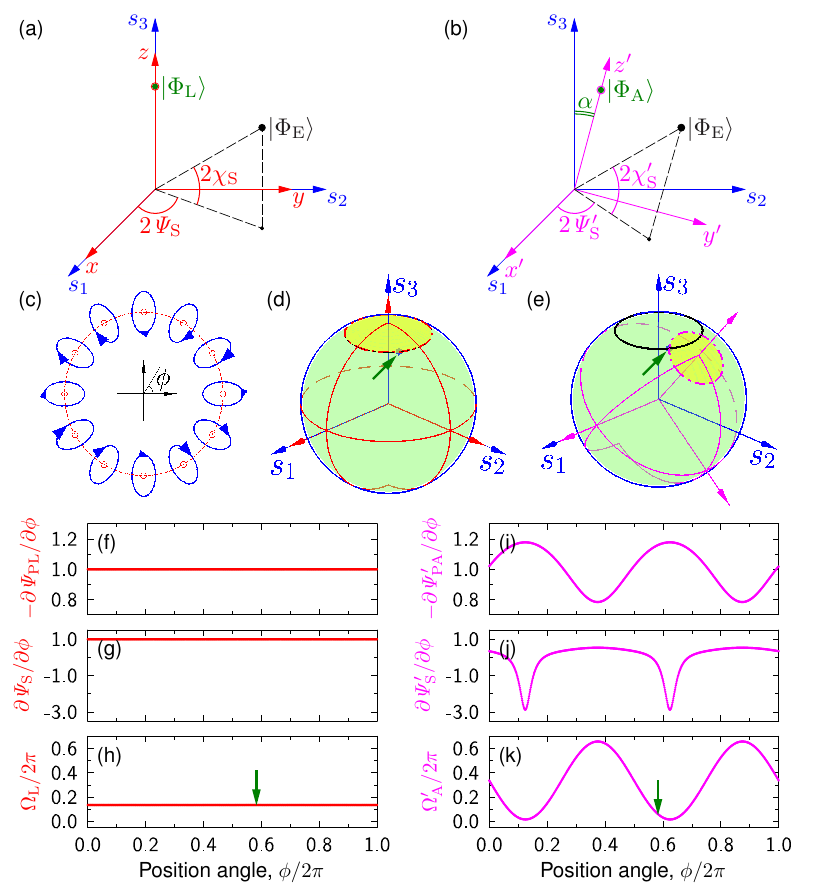} \caption{(a) Coordinate of the Poincar\'{e} sphere (blue axes) is represented
by $S_{1}-S_{2}-S_{3}$. Coordinate (red axes), represented by $x-y-z$
and used in the calculations, is contracted with reference field $\left|\Phi_{\text{L}}\right\rangle $
located at $z$ direction. Any studied field $\left|\Phi_{\text{E}}\right\rangle $
can be expressed as $\left(2\psi_{\text{S}},2\chi_{\text{S}}\right)$
in the coordinate of $x-y-z$. (b) Rotated coordinate (magenta axes),
represented by $x'-y'-z'$ and used in the calculations, is contracted
with reference field $\left|\Phi_{\text{A}}\right\rangle $ located
at $z$ direction. Any studied field $\left|\Phi_{\text{E}}\right\rangle $
can be expressed as $\left(2\psi'_{\text{S}},2\chi'_{\text{S}}\right)$.
(c) The field is simulated in (d-k). (d) and (e) SOP trace on the
Poincar\'{e} sphere, the yellow surface area corresponds to solid
angle of considered SOP marked by green arrow in calculation with
reference $\left|\Phi_{\text{L}}\right\rangle $ and $\left|\Phi_{\text{A}}\right\rangle $,
respectively. (f-h) Calculated three parts in equation \eqref{S23}.
(i-k) Calculated three parts in equation \eqref{S24}. }
\label{fig:S1} 
\end{figure*}

In the main text, the reference field was selected as right (left)
circularly polarized field in order to induce an elegant expression
(see equation \eqref{eq:OAM-charge} in the main text) for calculating
the OAM charge. But it does not mean that choice of general reference
fields would lead to any deviation. Here, we provide the calculation
process with simulation for the case of using general reference field.
For easy calculation, the selected reference field $\left|\Phi_{\text{A}}\right\rangle $
is located on the $S_{2}-S_{3}$ plane, and rotates an angle of $\alpha$
relative to $\left|\Phi_{\text{L}}\right\rangle $ around $S_{1}$
axis. Figure \ref{fig:S1}(a) and \ref{fig:S1}(b) illustrate coordinate
systems with the reference field of $\left|\Phi_{\text{L}}\right\rangle $
and $\left|\Phi_{\text{A}}\right\rangle $, respectively. In Fig.
\ref{fig:S1}(a), the contracted 3D coordinate (red axes of $x-y-z$)
holding $\left|\Phi_{\text{L}}\right\rangle $ located at $z$ direction
coincide with coordinate of the Poincar\'{e} sphere. Thus the investigated
vectorial vortex field $\left|\Phi_{\text{E}}\right\rangle $ in this
coordinate can be expressed as $\left(2\psi_{\text{S}},2\chi_{\text{S}}\right)$.
After coordinate transformation with assuming the new reference $\left|\Phi_{\text{A}}\right\rangle $
located at direction, as shown in Fig. \ref{fig:S1}(b), $\left|\Phi_{\text{E}}\right\rangle $
in the new contracted 3D coordinate (magenta axes of $x'-y'-z'$)
can be expressed as $\left(2\psi'_{\text{S}},2\chi'_{\text{S}}\right)$. 

Firstly, we consider the calculation of OAM charge with reference
$\left|\Phi_{\text{L}}\right\rangle $, combining equations \eqref{S18},
\eqref{S21}, \eqref{S22b}, and \eqref{S9b}, we rewrite the OAM
charge with the reference field $\left|\Phi_{\text{L}}\right\rangle $,
it reads 
\begin{equation}
l=\frac{\iint I_{\text{E}}\left(-\frac{\partial\psi_{\text{PL}}}{\partial\phi}+\frac{\partial\psi_{\text{S}}}{\partial\phi}\frac{\Omega_{\text{L}}}{2\pi}\right)rdrd\phi}{\iint I_{\text{E}}S_{0}rdrd\phi}.\label{S23}
\end{equation}
In equation \eqref{S23}, there are three parts in numerator: azimuthal
gradient of Pancharatnam phase ($-\partial\psi_{\text{PL}}/\partial\phi$),
azimuthal gradient of SOP-related phase ($\partial\psi_{\text{S}}/\partial\phi$),
and the relevant solid angle ($\Omega_{\text{L}}/2\pi$), which is
formed by the swept surface area of SOP revolving around the $\left|\Phi_{\text{L}}\right\rangle $
on the Poincar\'{e} sphere. With similar method, the calculation equation
of OAM charge under reference field $\left|\Phi_{\text{A}}\right\rangle $
can be written as 
\begin{equation}
l=\frac{\iint I_{\text{E}}\left(-\frac{\partial\psi_{\text{PA}}}{\partial\phi}+\frac{\partial\psi'_{\text{S}}}{\partial\phi}\frac{\Omega'_{\text{A}}}{2\pi}\right)rdrd\phi}{\iint I_{\text{E}}S_{0}rdrd\phi}.\label{S24}
\end{equation}
where $\psi_{\text{PA}}=\left\langle \Phi_{\text{A}}\right.\left|\Phi_{\text{E}}\right\rangle $
is Pancharatnam phase referenced to $\left|\Phi_{\text{A}}\right\rangle $,
$\psi'_{\text{S}}$ is the angle under new coordinate (see Fig. \ref{fig:S1}(b)),
and $\Omega'_{\text{A}}$ is solid angle of SOP swept area around
new reference $\left|\Phi_{\text{A}}\right\rangle $.

In order to verify the correctness of the equation \eqref{S24}, simulations
are carried out. Here, we consider the field shown in Fig.~\ref{fig:S1}(c)
(the same field shown in Fig.~\ref{fig1}(a) in main text), and calculate
the OAM charge with reference fields of $\left|\Phi_{\text{L}}\right\rangle $
and $\left|\Phi_{\text{A}}\right\rangle $, respectively. Figures
\ref{fig:S1}(f-h) show the calculated three parts in equation \eqref{S23}
with reference $\left|\Phi_{\text{L}}\right\rangle $, while Figs.~\ref{fig:S1}(i-k)
show the calculated three parts in equation \eqref{S24} with reference
$\left|\Phi_{\text{A}}\right\rangle $. In particular, the surface
areas of solid angle are illustrated Figs.~\ref{fig:S1}(d, e) for
a typical SOP, which is also marked in Figs.~\ref{fig:S1}(d, e,
h, k) with green arrows. Although each of three parts is different,
the calculated OAM charges of the light beam are equal with two reference
fields. From the above simulations, we believe that OAM charge can
be calculated by adopting any reference fields. However, with reference
field of right- or left-handed circular polarization, an elegant form
(equation \eqref{eq:OAM-charge} in main text) could be obtained.
Moreover, right- or left-handed circular polarization is the natural
choice to measure Stokes parameters. Thus, the reference field of
right- or left-handed circular polarization is adopted in the main
text.
\end{document}